\documentclass[submission,copyright,creativecommons]{eptcs}
\providecommand{\event}{HSB 2012} 
\usepackage{breakurl}             

\usepackage{ifpdf} 
\ifpdf 
\usepackage{graphicx}

\DeclareGraphicsExtensions{.pdf,.png,.mps}
\usepackage{pgf}
\else
\usepackage[dvipdfm]{graphicx}
\fi 

\usepackage[cmex10]{amsmath}
\usepackage{amsfonts,amsthm,amssymb}
\usepackage{stmaryrd}
\usepackage[mathscr]{euscript}
\usepackage{pgf}
\usepackage{tikz}
\usepackage{subfigure}
\usepackage{algorithm,algorithmic} 
\usepackage{enumitem}

\usetikzlibrary{arrows,automata}

\title{Hybrid Automata and $\epsilon$-Analysis on a Neural Oscillator\thanks{
This work has been partially supported by Istituto Nazionale di Alta Matematica (INdAM).}}
\author{Alberto Casagrande
\institute{Dept. of Matematics and Geoscience\\
University of Trieste, Italy}
\email{acasagrande@units.it}
\and
Tommaso Dreossi \qquad\qquad Carla Piazza
\institute{Dept. of Matematics and Computer Science\\
University of Udine, Italy}
\email{\quad dreossi.tommaso@spes.uniud.it \quad\qquad carla.piazza@uniud.it}
}
\def\titlerunning{Hybrid Automata and $\epsilon$-Analysis on a Neural Oscillator}
\def\authorrunning{A.~Casagrande, T.~Dreossi \& C.~Piazza}

\graphicspath{{./figures/}}

\newtheorem{definition}{Definition}
\newtheorem{example}{Example}
\newtheorem{theorem}{Theorem}

\def\bbbr{\mathbb{R}} 
\def\bbbn{\mathbb{N}} 

\newcommand{\Theory}{\mathcal{T}}
\newcommand{\ModelT}{\mathcal{M}}

\newcommand*{\llangle}{\left\langle\kern-2\nulldelimiterspace\left
\langle}
\newcommand*{\rrangle}{\right\rangle\kern-2\nulldelimiterspace\right
\rangle}

\newcommand{\VarSub}[3]{{#1}\llbracket{#2}/{#3}\rrbracket}
\newcommand{\VarSSub}[2]{{#1}\llbracket{#2}\rrbracket}
\newcommand{\NotBoundIn}[2]{{#1}[{#2}]}

\newcommand{\formul}{\varphi}

\newcommand{\limply}{\rightarrowtriangle}

\newcommand{\VarNotation}[3]{{#1}%
\ifthenelse{\equal{#3}{}}{%
\ifthenelse{\equal{#2}{}}%
{}%
{_{#2}}}%
{_{({#3}){#2}}}}

\newcommand{\Xname}{X} 
\newcommand{\X}[2][]{\VarNotation{\Xname}{#2}{#1}}   
\newcommand{\vX}[1]{\mathbf{\X{#1}}}

\newcommand{\Yname}{Y} 
\newcommand{\Y}[2][]{\VarNotation{\Yname}{#2}{#1}}   
\newcommand{\vY}[1]{\mathbf{\Y{#1}}}

\newcommand{\Wname}{W} 
\newcommand{\W}[2][]{\VarNotation{\Wname}{#2}{#1}}   
\newcommand{\vW}[1]{\mathbf{\W{#1}}}

\newcommand{\Loc}{\mathscr{V}}
\newcommand{\Edg}{\mathscr{E}}

\newcommand{\ActSymb}{\mathit{Act}}
\newcommand{\ResetSymb}{\mathit{Res}}
\newcommand{\DynSymb}{\mathit{Dyn}}
\newcommand{\InvSymb}{\mathit{Inv}}

\newcommand{\Tvar}[1]{T_{#1}}

\newcommand{\ReachSet}[3][]{{\mathit{RSet}_{#2}^{#1}\!\left({#3}\right)}}

\newcommand{\Act}[2][]{\ActSymb_{#1}{(#2)}}
\newcommand{\ActDef}[2][]{\NotBoundIn{\Act[#1]{#2}}{\vX{}}}
\newcommand{\ActVal}[3][]{\VarSSub{\Act[#1]{#2}}{#3}}

\newcommand{\HReset}[2][]{\ResetSymb_{#1}({#2})}
\newcommand{\HResetDef}[2][]{\NotBoundIn{\HReset[#1]{#2}}{\vX{},\vX{}'}}
\newcommand{\HResetVal}[3][]{\VarSSub{\HReset[#1]{#2}}{#3}}

\newcommand{\Inv}[2][]{\InvSymb_{#1}({#2})}
\newcommand{\InvDef}[2][]{\NotBoundIn{\Inv[#1]{#2}}{\vX{}}}
\newcommand{\InvVal}[3][]{\VarSSub{\Inv[#1]{#2}}{#3}}

\newcommand{\Dyn}[2][]{\DynSymb_{#1}({#2})}
\newcommand{\DynDef}[2][]{\NotBoundIn{\Dyn[#1]{#2}}{\vX{},\vX{}',T}}

\newcommand{\vfFunct}[1]{{f}_{#1}}

\newcommand{\obmap}[1]{\llangle{#1}\rrangle}

\newcommand{\trans}[2][]{\xrightarrow{#2}_{#1}}
\newcommand{\ctrans}[1]{\trans[C]{#1}}
\newcommand{\dtrans}[1]{\trans[D]{#1}}
\newcommand{\actrans}{\rightarrow_{C}}
\newcommand{\adtrans}{\rightarrow_{D}}

\newcommand{\defeq}{\stackrel{\text{\tiny def}}{=}}

\newcommand{\tuple}[1]{\langle{#1}\rangle}

\newcommand{\state}[2]{\left\langle{{#1},{#2}}\right\rangle}

\newcommand{\dims}[1]{d({#1})}

\newcommand{\dsem}[2][]{\left\{\kern-2\nulldelimiterspace\left|{#2}\right|\kern-2\nulldelimiterspace\right\}_{#1}}

\newcommand{\sphseml}{\left(\kern-2\nulldelimiterspace\left|}
\newcommand{\sphsemr}{\right|\kern-2\nulldelimiterspace\right)}
\newcommand{\sphsem}[2][\epsilon]{\sphseml{#2}\sphsemr_{#1}}

\newcommand{\seml}{\left\{\kern-2\nulldelimiterspace\left|}
\newcommand{\semr}{\right|\kern-2\nulldelimiterspace\right\}}
\newcommand{\epssem}[2][\epsilon]{\seml{#2}\semr_{#1}}
\newcommand{\stdsem}[1]{\seml{#1}\semr}

\newcommand{\ball}[2]{B\left({#1},{#2}\right)}
\newcommand{\eball}[1]{\ball{#1}{\epsilon}}

\DeclareMathAlphabet{\mathpzc}{OT1}{pzc}{m}{it}

\newcommand{\sgn}{\operatorname{sgn}}

\newcommand{\ath}[2][\lambda,\alpha]{h_{#1}({#2})}

\usepackage{xifthen}

\newcommand{\hstate}[1]{%
\ifthenelse{\isempty{#1}}%
    {a}
    {a_{#1}}
}

\newcommand{\hloc}[1]{v_{#1}}

\def\edist{\delta}

\newcommand{\relation}{\fallingdotseq}
\newcommand{\mult}{*}

\newcommand{\tostdsem}[2][\epsilon]{\widehat{\left(#2\right)}_{#1}}

\begin{document}
\maketitle

\begin{abstract}
In this paper we propose a hybrid model of a neural oscillator, obtained by partially discretizing a well-known continuous model.
Our construction points out that in this case the standard techniques, based on replacing sigmoids with step functions, is not 
satisfactory.
Then, we study the hybrid model through both symbolic methods and approximation techniques. This last analysis, in particular, allows
us to show the differences between the considered approximation approaches.
Finally, we focus on approximations via $\epsilon$-semantics, proving how these can be computed in practice. 
\end{abstract}

\section*{Introduction}\label{sec:intro}


Neural oscillations are rhythmic and repetitive electrical stimuli which play an important role in 
the activities of several brain regions. Some examples of brain locations in which it has been demonstrated 
the central role of neural oscillations are 
the cortex~\cite{1824881}, the thalamus~\cite{citeulike:4043128}, and the olfactory information processing~\cite{citeulike:762104}. 
With the aim of understanding neurophysiological activities, we propose the modeling 
of oscillatory phenomena exploiting hybrid automata. 

A continuous model of a single oscillator based on an ordinary differential system has been proposed
in~\cite{Tonnelier19991213}. Even if this is a simple model, its analysis and the analysis of its composition in multiple
copies is limited due to the non-linearity of the ordinary differential system involved. For this reason, 
we are interested in the development of a piecewise affine hybrid automaton which correctly approximates 
the continuous model and on which automatic analysis and composition can be made.

Trying to linearize the non-linear components of the original continuous model, we first replace in the 
standard way the sigmoidal behaviours with sign functions~\cite{tomlinmodel,springerlink:10.1016/j.bulm.2003.08.010}, approximating 
continuos signals with discrete off-on signals. Unfortunately, the behaviour of this model differs from the original one. For 
this reason, we propose a more sophisticated approximation of sigmoidals based on a piecewise 
linear function exploited in the development of a hybrid automaton which simulates a single oscillator.

It is well known that the reachability problem over hybrid automata
is source of undecidability.
Moreover, the exact computation of the reachable sets of hybrid automata, which represents the basis of 
the automatic analysis of the models, does not always reflect the behaviour of the real modeled systems. This is 
due to the fact that real systems are often subject to noise, thus their evolutions do not correspond to 
a single precise formalization. This has been noticed already in~\cite{Tonnelier19991213} in the specific case of the continuous model of 
the neural oscillator. For these reasons, we study our hybrid automaton exploiting different 
approximation techniques that introduce noise. 

In the literature, several approximation techniques have been proposed (see, e.g.,~\cite{franzle99,DBLP:conf/tamc/Ratschan10,Girard:2007fk,
DBLP:journals/deds/CasagrandePP09,Prabhakar:2009:VTS:1683310.1684907}). 
Fr\"anzle in~\cite{franzle99}
presents a model of noise over hybrid automata. The introduction of noise ensures
in many cases the (semi-)decidability of the reachability problem. Another result of 
(semi-)decidability always based on the concept of perturbation and concerning the safety
verification of hybrid systems is given by Ratschan in~\cite{DBLP:conf/tamc/Ratschan10}. Furthermore,
$\epsilon$-(bi)simulation~\cite{Girard:2007fk}
relations, which are essentially relaxations on the infinite precision required by simulation and bisimulation, 
represent tools able to remove complexity and undecidability issues related to the analysis of the
investigated model. Moreover, in~\cite{DBLP:journals/deds/CasagrandePP09} it is presented a different approach based
on the reinterpretation of the standard semantics of the formul\ae~which compose hybrid automata.
Exploiting this new class of semantics, called $\epsilon$-semantics, the authors provide a 
result of decidability of the reachability problem over hybrid automata with bounded invariants.

In this paper, we focus precisely on the approximation approach based on the $\epsilon$-semantics.
In particular, we propose a translation that allows us to reduce the  
$\epsilon$-semantics evaluation to the standard semantics evaluation, computable by exploiting 
tools for cylindrical algebraic decomposition. Then, we present some properties which have been automatically 
tested on the neural oscillator by applying such translation. Hence, in this work, we prove both that 
$\epsilon$-semantics better represents the real behaviour of the neural oscillator, than the standard one, 
and that the approach is effective. 

The paper is organized as follows: Section~\ref{sec:ha} gives some basic definitions concerning logics and
hybrid automata; Section~\ref{sec:model} is dedicated to the mathematical modeling of the neural oscillator;
in Section~\ref{sec:approx} we present different approximation techniques based on noise, perturbation,
approximate (bi)simulations, and $\epsilon$-semantics. Section~\ref{sec:analysis} exposes some considerations regarding 
the application of the previously presented  approximation approaches to the investigated model.
Finally, in Section~\ref{sec:computing}, we first define a translation which make effectively computable the
$\epsilon$-semantics, then we experimentally exploit it in the analysis of the hybrid automaton
which models the neural oscillator. All the proofs can be found at \url{http://www.dimi.uniud.it/piazza/hsb2012_extended.pdf}.

\section{Hybrid Automata}\label{sec:ha}
\subsection{Preliminaries}\label{sec:logics}

We formally define hybrid automata by using first-order languages and, 
because of that, we first need to introduce some basic notions and our notation.

We use $\X{}$, $\X{i}$, $\Y{}$,  $\Y{i}$,  $\W{}$,  and $\W{i}$ to denote real variables and 
 $\vX{}$, $\vX{i}$, $\vY{}$,  $\vY{i}$,  $\vW{}$,  and $\vW{i}$ to denote tuple 
 of real variables. 
We always assume that all the variables that occur bound in a formula do
not occur free and vice versa. This enables us to label variables, rather than occurrences,  
as free or bound.
We write 
$\NotBoundIn{\formul}{\X{1}, \dots, \X{m}}$ to 
stress the fact that 
$\X{1}$, $\ldots$, $\X{m}$ are free in $\formul$.
By extension, 
$\NotBoundIn{\formul}{\vX{1}, \ldots, \vX{n}}$ 
indicates that  
 the components of vectors 
$\vX{1}$, $\ldots$, $\vX{n}$ are free in $\formul$.

%
The formula obtained from $\NotBoundIn{\formul}{\X{1}, \ldots,\X{m}}$ by replacing $\X{i}$ by $s_0$, 
where $s_0$ is either a constant or a variable,  is
denoted by $\VarSub{\formul}{\X{i}}{s_0}$.
By extension, $\VarSub{\formul}{\X{i}\ldots\X{i+n}}{s_{0}\ldots s_{n}}$ 
indicates the formula obtained from $\NotBoundIn{\formul}{\X{1}, \ldots,\X{m}}$
by simultaneously replacing all the variables $\X{i}\ldots\X{i+n}$ by $s_{0}\ldots s_{n}$. 
If $\vX{}=\tuple{\X{i},\ldots,\X{i+n}}$, $\vec{s_0}=\tuple{s_{0},\ldots,s_{n}}$, 
$\vec{s_1}=\tuple{s_{n+1},\ldots,s_{2\mult{}n+1}}$, and 
$\relation$ is a relational symbol (e.g., $=$ or $\geq$), then we may write 
$\VarSub{\formul}{\vX{}}{\vec{s_0}}$ in place of 
$\VarSub{\formul}{\X{i}\ldots\X{i+n}}{s_{0}\ldots s_{n}}$ and 
$\vec{s_0}\relation\vec{s_1}$ in place of 
$\bigwedge_{j \in [0,n]} \left( {s_{j}\relation s_{j+k\mult{}n}} \right)$ (e.g., 
$\tuple{7,\X{}} = \tuple{2,3}$ means $7=2 \land \X{}=3$). 
Finally, if $\NotBoundIn{\formul}{\vX{1}, \ldots, \vX{i}, \ldots, \vX{n}}$ then we may denote the 
formula $\VarSub{\formul}{\vX{i}}{\vec{s}}$ by writing  
$\VarSSub{\formul}{\vX{1}, \ldots, \vec{s}, \ldots, \vX{n}}$.

The semantics of a formula 
is defined in the standard way (see. e.g.,~\cite{ender}).
%
Given a set $\Gamma$ of sentences and a sentence $\formul$, we say
that $\formul$ is a \emph{logical consequence} of $\Gamma$ (denoted,
$\Gamma\models\formul$) if 
$\formul$ is valid in any model $\ModelT$  in which each formula of $\Gamma$ is valid  too
($\ModelT \models \Gamma$).
A \emph{theory} $\Theory$ is a set of sentences such that if
$\Theory\models \formul$, then $\formul\in\Theory$.
A theory $\Theory$
admits the so-called \emph{elimination of quantifiers}, if, for any
formula $\formul$, there exists in $\Theory$ a quantifier free formula $\varrho$  
such that $\formul$ is equivalent to $\varrho$ with respect to
$\Theory$. If there exists an algorithm for deciding whether a
sentence $\formul$ belongs to $\Theory$ or not, we say that
$\Theory$ is \emph{decidable}. 

\begin{example}
Consider the formula 
$\formul \defeq \exists \X{}\, (a \mult \X{}^2 + b\mult \X{} + c = 0)$. 
It is well known that $\formul$ is in the theory of reals with $+$, $\mult{}$, 
and $\geq$ if and only if the unquantified formula $b^2 - 4 ac \geq 0$ holds.
\end{example}

%

\subsection{Syntax, Semantics, and Reachability}\label{sec:hybrid}

%

A hybrid automaton is an infinite state automaton that consists in a 
set of continuos variables and a finite directed graph. 
Each node of a graph is labelled by both an invariant condition and a
dynamic law, while all the edges are tagged with an activation region and 
a reset map. 
The continuous variables evolve according to the dynamic law 
of the current node of the graph and the node's 
invariant condition must be satisfied along all the evolution. 
An edge is crossable if and only if the variable values are included the 
activation region and, when a hybrid automata jumps over it, 
the associated reset map is applied. 

\begin{definition}[Hybrid Automata - Syntax]\label{def:hyau}
   A \emph{hybrid automaton} $H$ of dimension $\dims{H} \in \bbbn$ is a tuple 
   $\langle\vX{}{}$, $\vX{}'$, $\Loc$, $\Edg$, $\InvSymb$,
   $\vfFunct{\cdot}$,  $\ActSymb$, $\ResetSymb\rangle$ where:
   \begin{itemize}
   \item $\vX{}{}=\langle\X{1}$, $\ldots$, $\X{n}\rangle$ and
   $\vX{}'=\langle\X{1}'$, $\ldots$, $\X{n}'\rangle$ are two vectors of
   variables ranging over the reals $\bbbr$;
   \item $\langle \Loc$,  $\Edg\rangle$ is a directed finite graph, i.e., $\Edg\subseteq \Loc\times \Loc$. Each element
   of $\Loc$ will be dubbed \emph{location};
   \item Each location $v\in \Loc$ is labelled by both a formula
   $\InvDef{\hloc{}}$, called \emph{invariant}, and a continuous function 
   $\vfFunct{\hloc{}}: \bbbr^{n} \longrightarrow (\bbbr_{\geq 0} \longrightarrow \bbbr^{n})$, 
   called  \emph{dynamics} or \emph{flow function}. The dynamics may be specified either by differential equations, i.e., $\vfFunct{\hloc{}}$ is the solution 
of a given Cauchy problem, or by a logic formula. We use the formula 
$\DynDef{\hloc{}}$, where $T$ is a temporal variable ranging in $\bbbr_{\geq 0}$, to denote the dynamics on
$v$, i.e., $\DynDef{\hloc{}}\defeq \vX{}'=\vfFunct{\hloc{}}(\vX{}{})(T)$;
   \item Each $e\in \Edg$ is labelled by the formul\ae\
   $\ActDef{e}$ and $\HResetDef{e}$ which are
   called \emph{activation} and \emph{reset}, respectively.
\end{itemize}
\end{definition}


%
If 
all the formul\ae\ that define a hybrid automaton $H$ belong
to
the same logical theory $\Theory$, then
we say 
that $H$ is \emph{definable in $\Theory$} or that $H$ is a
\emph{$\Theory$ hybrid automaton}.


The semantics of any hybrid automaton can be specified as a transition system that 
 is composed by two different relations 
miming the double nature of the hybrid automaton itself: the 
\emph{continuous reachability transition relation} and the \emph{discrete reachability transition 
relation}.

\begin{definition}[Hybrid Automaton - Semantics]\label{def:semantics}
A \emph{state} $\hstate{}$ of $H$ is a pair $\state{\hloc{}}{r}$, where $v\in \Loc$
is a location and $r\in \bbbr^{\dims{H}}$ 
is an assignment of values for the variables of $\vX{}{}$.
A state $\state{\hloc{}}{r}$ is said to be \emph{admissible} if
$\InvVal{\hloc{}}{r}$ holds.

The \emph{continuous transition relation} $\ctrans{t}$ between admissible states, 
where $t\geq 0$ denotes the transition elapsed time, is defined as follows:

\vskip0.0cm
\noindent
\begin{tabular}{lcl}
$\state{\hloc{}}{r} \ctrans{t} \state{\hloc{}}{s}$& $\iff$  &
$r=\vfFunct{\hloc{}}(r)(0)$, 
$s=\vfFunct{\hloc{}}(r)(t)$, and 
$\InvVal{\hloc{}}{\vfFunct{\hloc{}}(r)(t')}$ hold for each  $t'\in[0,t]$. 
\end{tabular}

The \emph{discrete transition relation} $\dtrans{e}$ among
admissible states is:

\vskip0.15cm
\noindent
\begin{tabular}{lcl}
$\state{\hloc{}}{r} \dtrans{e} \state{\hloc{}}{s}$& $\iff$  &
$e=\tuple{\hloc{},\hloc{}'}$ 
and both 
$\ActVal{e}{r}$ and $\HResetVal{e}{r,s}$ hold.
\end{tabular}
\end{definition}
We write $\hstate{}\actrans \hstate{}'$ and $\hstate{}\adtrans \hstate{}'$ to mean 
that there exists a $t \in \bbbr_{\geq0}$ such that
$\hstate{}\ctrans{t} \hstate{}'$ and that there exists an $e \in \Edg$ such that
$\hstate{}\dtrans{e} \hstate{}'$, respectively.


%

\begin{definition}[Reachability]\label{semantics2}
Let $\mathcal{I}$ be either $\bbbn$ or an initial finite interval of $\bbbn$.
A \emph{trace} of $H$ is a sequence of admissible states $\hstate{0},\hstate{1},\dots,
\hstate{j},\dots$, with $j\in \mathcal{I}$, such that
$\hstate{i-1}\rightarrow \hstate{i}$ holds for all $i \in [1,n]$ and 
either $\hstate{i-2}\actrans \hstate{i-1}\adtrans\hstate{i}$,
$\hstate{i-2}\adtrans \hstate{i-1}\adtrans\hstate{i}$,
or $\hstate{i-2}\adtrans \hstate{i-1}\actrans\hstate{i}$ for each $i\in I\setminus \{0\}$\footnote{This last condition 
supports not transitive dynamics. See~\cite{focore2008} for a complete discussion.}.

The automaton $H$ \emph{reaches} a state $\hstate{n}$ from a state 
$\hstate{0}$ if
there exists a trace $\hstate{0},\dots,\hstate{n}$.
In such a case, we also say that $\hstate{n}$ is \emph{reachable} from $\hstate{0}$ in $H$.
\end{definition}

The problem of deciding whether a hybrid automaton $H$ 
reaches a set of states $S$ from a set of states $R$ is known
as the \emph{reachability problem} of $S$ from $R$ over $H$. 
A trace produced by an infinite sequence of discrete transitions during
a bounded amount of time is called \emph{Zeno trace} and
every hybrid automaton allowing such kind of trace is said to have a
\emph{Zeno behaviour}.

\begin{example}\label{example:ball}
Let us consider a hybrid automaton $H_b$ modeling a
\emph{bouncing ball} whose collisions are inelastic.
\begin{figure}[!h]
\begin{center}
\begin{tikzpicture}[->,>=stealth',shorten >=1pt,auto,node distance=2.8cm,
                    semithick,scale=0.75, transform shape]
  \tikzstyle{every state}=[fill=blue!20!white,draw,text=black]

  \node[state]   (S0)  {\begin{tabular}{c}
                                   $\dot{X}_1=\X{2}$ \\ 
                                   $\dot{X}_2=-g$
                                   \end{tabular}};

  \path (S0) edge [loop right] node {
                                   \begin{tabular}{l}
                                   $\X{1}'=\X{1}$ \\ 
                                   $\X{2}'=-\gamma \X{2}$
                                   \end{tabular}} (S0);
\end{tikzpicture}
\caption{Bouncing ball hybrid automaton.}\label{fig:bouncing}
\end{center}
\end{figure}
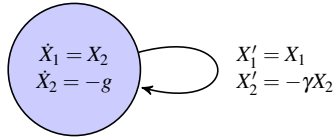
The automaton is equipped with two continuous variables $\X{1}$ and $\X{2}$ 
that represent ball's elevation and 
velocity, respectively.
The dynamics, resets, and discrete structure of
$H_b$ are presented in Fig.~\ref{fig:bouncing}.
The two coefficients $g$ and $\gamma$ are the standard gravity
and the coefficient of restitution, respectively. The activation
formula of the automaton edge is ``$\X{1} = 0$''.

Imposing as starting height $h_0=10m$ and as  coefficient of restitution $\gamma=0.86$, 
the 
bounce peaks decrease 
at each iteration and 
the automaton $H_b$ has a Zeno behaviour. 
%
\end{example}

As the halting problem for the two counter machine can be reduced to 
the reachability problem of a particular class of hybrid automata, 
the reachability problem for hybrid automata itself is not always decidable \cite{alur95}. 
However, if $H$ is a $\Theory$-hybrid automaton and $\Theory$ is a first-order 
decidable theory, then the reachability through a bounded number of discrete transitions 
can be characterized with a first-order decidable formula (see e.g.,~\cite{focore2008}). In particular, 
in the case of automata defined through polynomials over the reals, we can use 
cylindrical algebraic decomposition tools to decide bounded reachability.

\section{Neural Oscillator: Continuous and Hybrid Models}\label{sec:model}

Oscillatory electrical stimuli have been considered central for the activities 
of  several brain regions since the begin of the '80s. It was shown that they play 
an important role in the olfactory information processing~\cite{citeulike:762104} 
and they were observed 
in 
the thalamus~\cite{citeulike:4043128}, and in the cortex~\cite{1824881}. 
Many studies suggested that, in the mammalian visual system, 
neurons signals may be group together through in-phase 
oscillations~\cite{citeulike:547881}. 
Because of this, the development and analysis of models representing 
oscillatory phenomena assume a great importance in 
understanding  the neurophysiological activities.

\begin{figure}[!ht]
 \centering
 \subfigure[An evolution of $f(3,1)$.]
   {\includegraphics[width=0.48\textwidth]{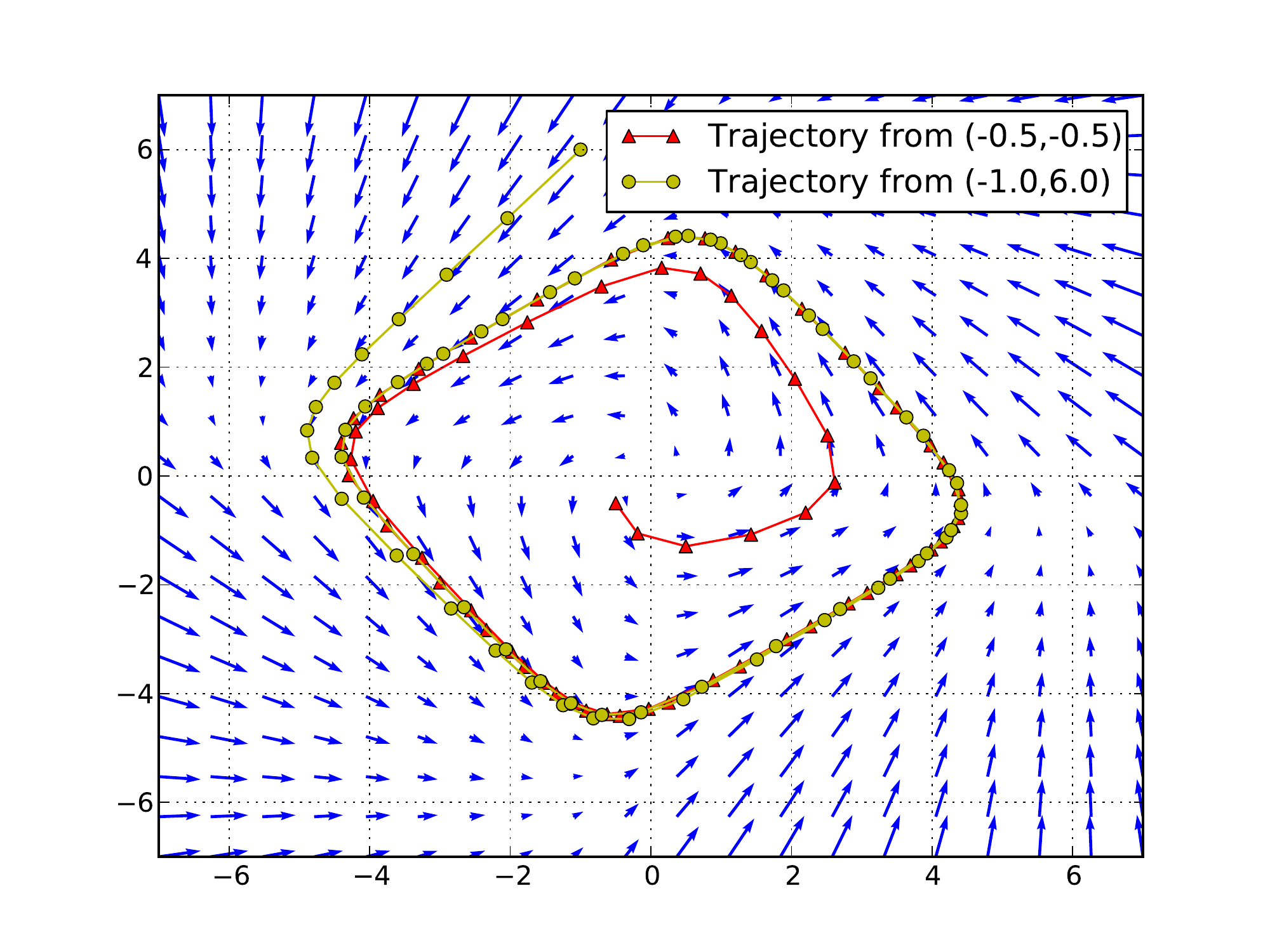}\label{fig:origevo}}
 \hspace{3mm}
 \subfigure[An evolution of $\tilde{f}_{2}(3,1)$.]
   {\includegraphics[width=0.48\textwidth]{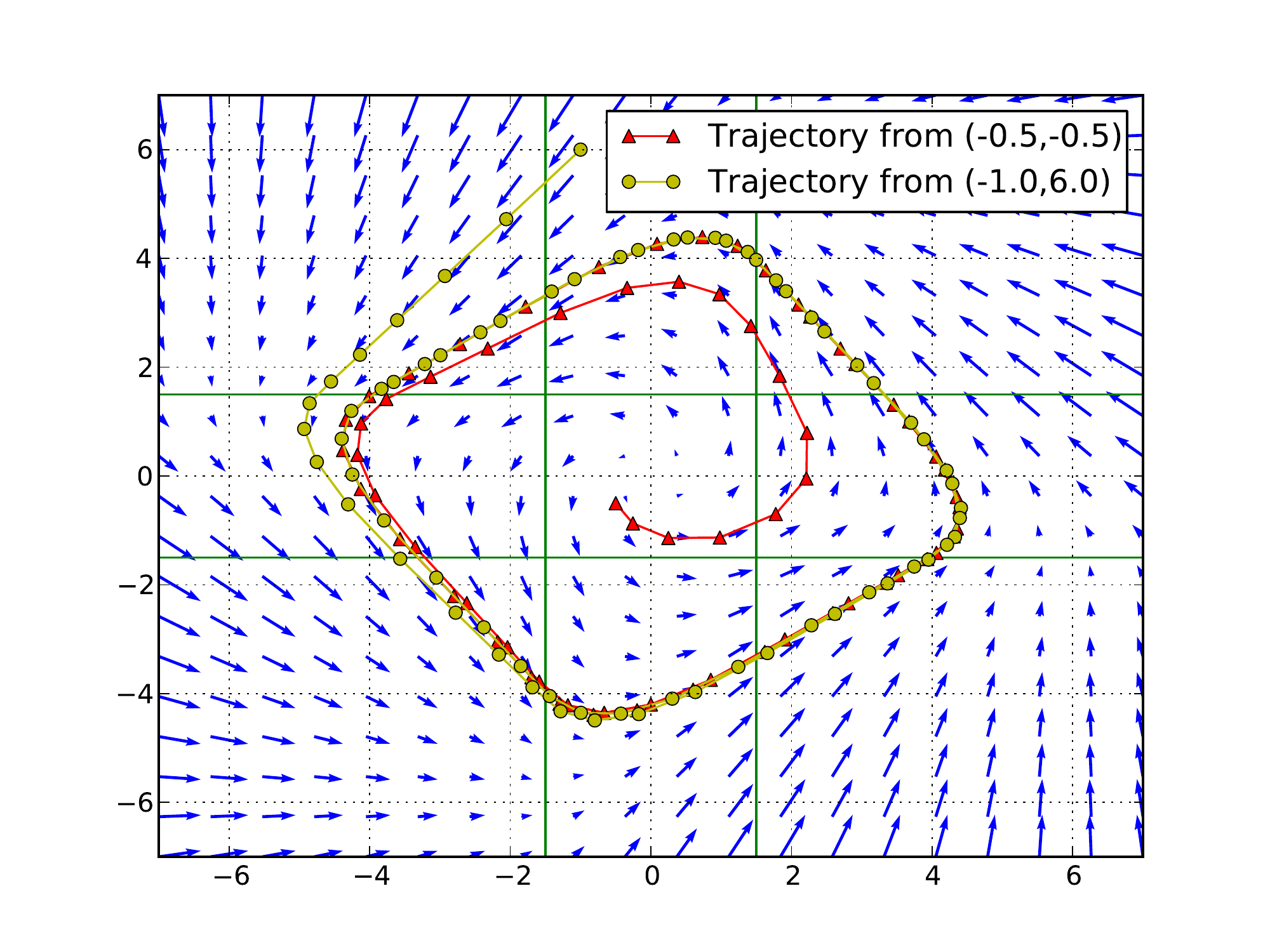}\label{fig:goodevo}}
    \caption{Direction field and evolution of the models discussed in Section~\ref{sec:model}.}\label{fig:diamond_cycle}
 \end{figure}

A simple continuous model of a single oscillator has been proposed in~\cite{Tonnelier19991213}. 
The model describes the evolutions of one excitatory neuron ($N_e$) and one inhibitory neuron ($N_i$) 
by mean of the ordinary differential system.  
\begin{equation}\label{eq:origsystem}
f(\tau,\lambda):\; \left\{\begin{array}{l}
\dot{\X{e}}=-\frac{\X{e}}{\tau}+\tanh{(\lambda*\X{e})}-\tanh{(\lambda*\X{i})}\\
\dot{\X{i}}=-\frac{\X{i}}{\tau}+\tanh{(\lambda*\X{e})}+\tanh{(\lambda*\X{i})}
\end{array}\right., 
\end{equation}
where $\X{e}$ and $\X{i}$ are the output of $N_e$ and $N_i$, respectively, 
$\tau$ is a characteristic time constant, and $\lambda>0$ is the amplification gain.
Hopf bifurcation characterizes a qualitative change in the evolution of $f(\tau,\lambda)$: 
if $\tau*\lambda\leq 1$, then the point $\tuple{0,0}$ is the unique global attractor of the system, if, otherwise, 
$\tau*\lambda>1$, the origin is an unstable equilibrium and all the evolutions converge to a limit cycle 
attractor
~\cite{DBLP:journals/ijns/AtiyaB89}.
A simulation of $f(3,1)$ is represented in Fig.~\ref{fig:origevo}.

\begin{figure}[!h]
\begin{center}
 \subfigure[The automaton 
  has 4 locations. 
The dashed lines denote both the boundaries of the invariants and the activation regions.  
The resets are identify functions.]
   {\includegraphics[width=0.41\textwidth]{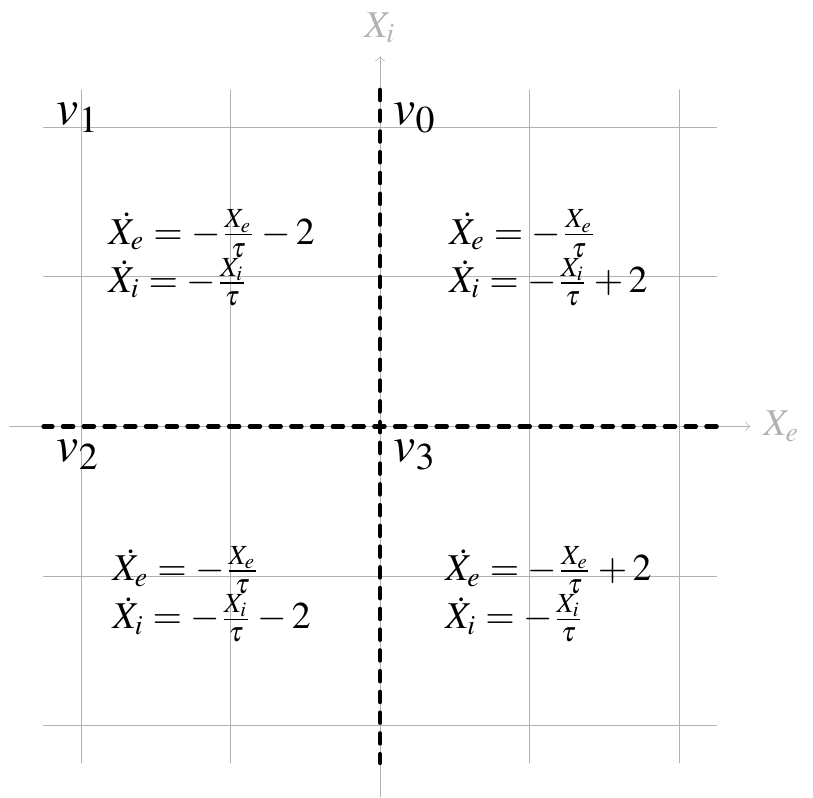}\label{fig:badapprox}}
    \hspace{5mm}
 \subfigure[Direction field and evolution of the automaton  
 from the two points $\tuple{-\frac{1}{2},-\frac{1}{2}}$ and 
 $\tuple{-1,6}$ when $\tau=3$. The automaton has four attractors, is not periodic, and   
its principal axes are stable.]
   {\includegraphics[width=0.50\textwidth]{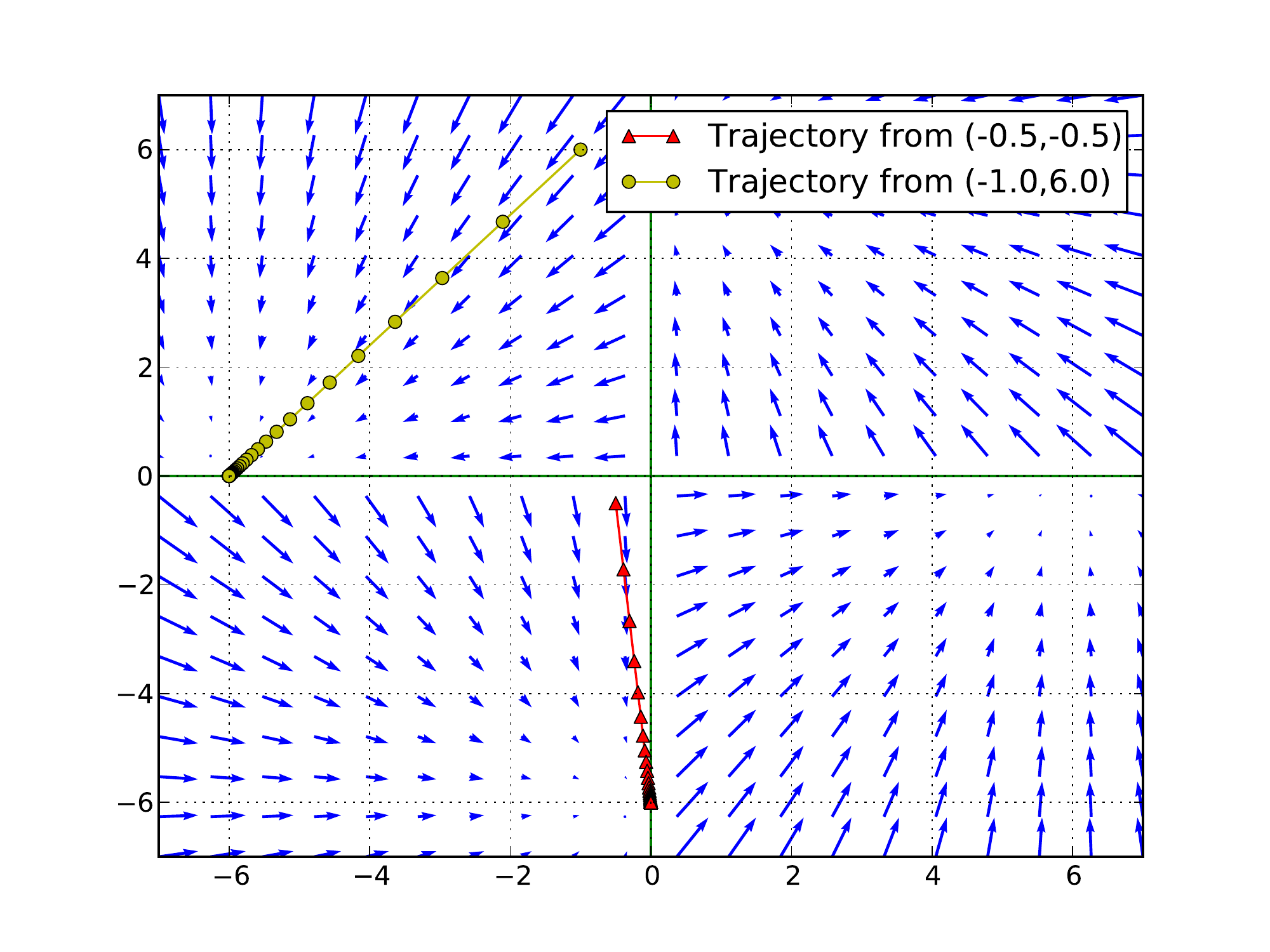}\label{fig:badevo}}

\caption{The piecewise hybrid automaton associated to the function $\hat{f}(\tau)$.%
}
\end{center}
\end{figure}

Even if $f(\tau,\lambda)$ 
is rather simple, the ability of analyzing a complex system obtained 
by composing multiple copies of this model is limited due to the non-linearity of $f(\tau,\lambda)$ itself. 
For this reason, we are interested in the development of a piecewise affine 
hybrid model whose behaviour fairly approximates System (\ref{eq:origsystem}) and that can be automatically 
analyzed and composed. 

Since the non-linear components in System~(\ref{eq:origsystem}) have the form $\tanh{(\lambda*\X{})}$, 
we try to linearize such function. 
In the case of genetic networks it is quite standard to approximate sigmoidal behaviours (e.g., $\tanh$) through 
the sign function $\sgn$~\cite{tomlinmodel,springerlink:10.1016/j.bulm.2003.08.010}. Such approximation replaces a continuous signal with a discrete 
off-on one. 
In our case, by replacing $\tanh{(\lambda*\X{})}$ 
with $\sgn{(\X{})}$ in System~(\ref{eq:origsystem}), we obtain the following differential 
system:
\begin{equation}
\hat{f}(\tau):\; \left\{\begin{array}{l}
\dot{\X{e}}=-\frac{\X{e}}{\tau}+\sgn{(\X{e})}-\sgn{(\X{i})}\\
\dot{\X{i}}=-\frac{\X{i}}{\tau}+\sgn{(\X{e})}+\sgn{(\X{i})}
\end{array}\right. ,
\end{equation}
which corresponds to the piecewise hybrid model depicted in Fig.~\ref{fig:badapprox}.
Unfortunately, the behaviour of this model is quite different from that of System~(\ref{eq:origsystem}), as 
we can see comparing the simulation in Fig.~\ref{fig:badevo} with that of Fig.~\ref{fig:origevo}. In particular, 
the model based on $\hat{f}(\tau)$ has four 
attractors with coordinates $\tuple{-2*\tau,0}$, $\tuple{0,-2*\tau}$, $\tuple{2*\tau,0}$, and $\tuple{0,2*\tau}$, it is 
not periodic, and  its principal axes are stable.  

\begin{figure}[!h]
\begin{center}
\includegraphics[width=0.67\textwidth]{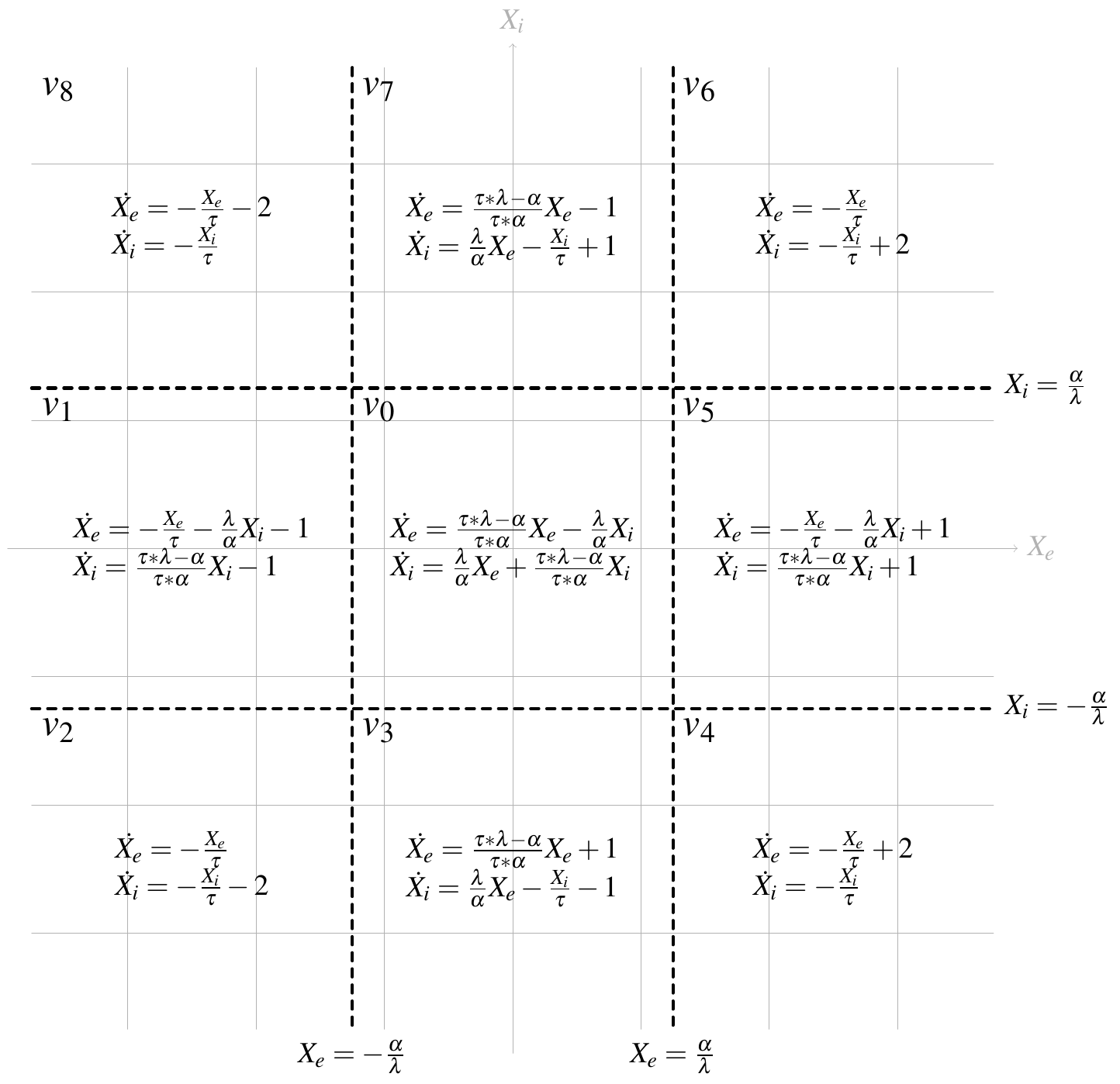}
\caption{A graphical representation of the hybrid automaton $H_{\tilde{f}}$ associated to the function $\tilde{f}_{\alpha}(\tau,\lambda)$.}\label{fig:goodapprox}
\end{center}
\end{figure}

A more sophisticated approximation of $\tanh{(\lambda*\X{})}$ 
is the piecewise linear function: 
\begin{equation}
\ath[{\lambda,\alpha}]{z} \defeq \left\{\begin{array}{l l}
-1 & \text{if }z< -\frac{\alpha}{\lambda}\\
\frac{\lambda}{\alpha}*z & \text{if }-\frac{\alpha}{\lambda} \leq z < \frac{\alpha}{\lambda}\\
1 & \text{if }z\geq \frac{\alpha}{\lambda}
\end{array}\right. ,
\end{equation}
where $\alpha$ is the approximation coefficient which determines the slope 
of the central segment. 
The substitution of $\tanh{(\lambda*\X{})}$ with 
$\ath[{\lambda,\alpha}]{z}$ in System~(\ref{eq:origsystem}) leads to 
the system:  
\begin{equation}
\tilde{f}_{\alpha}(\tau,\lambda):\; \left\{\begin{array}{l}
\dot{\X{e}}=-\frac{\X{e}}{\tau}+\ath{\X{e}}-\ath{\X{i}}\\
\dot{\X{i}}=-\frac{\X{i}}{\tau}+\ath{\X{e}}+\ath{\X{i}}
\end{array}\right. ,
\end{equation}
whose corresponding hybrid automaton  is depicted in Fig.~\ref{fig:goodapprox}. 
A different 
approximation could be obtained by using the technique 
proposed in~\cite{Grosu:2011fk}. 

 In the rest of the paper we present some general techniques for studying hybrid automata and
 then we apply them to $H_{\hat{f}}$ to formally prove its properties.

\section{Approximation Techniques}\label{sec:approx}
\subsection{Noise and Disturbed Automata}\label{sec:franzle}
The density of continuous variables provide an unbounded quantity of memory within a bounded region.
As a matter of fact, the undecidability results proved in \cite{undecidable} are based on 
the possibility of embedding $\mathbb{N}$ in $(0,1]\subseteq \bbbr$ through the function $f(n)=2^{-n}$.

However, Fr\"anzle in \cite{franzle99} observed that noise disturbes the trajectiories of real hybrid systems,
augmenting the set of reachable points. 
Hence, in \cite{franzle99} a model of noise has been presented over hybrid automata.
Remarkably, the introduction of noise ensures in many cases the (semi-)decidability of the reachability problem.

Our definitions of hybrid automata slightly differ from the ones in \cite{franzle99}. In particular, as far as the syntax is concerned, the formul\ae\ $\ActDef{e}$ and $\HResetDef{e}$ are glued together in a formula called $trans_{e}[\vX{},\vX{}']$. Moreover, our formul\ae\ $\InvDef{v}$ and $\DynDef{v}$ are replaced by a single formula $act_{v}[\vX{},\vX{}']$ whose meaning in our framework is:
$$\begin{array}{l}\exists T(T\geq 0\wedge \vX{}'=\vfFunct{v}(\vX{})(T)\wedge \forall T'(0\leq T'\leq T \rightarrow \Inv{v}{\vfFunct{v}(\vX{})(T')})), 
\end{array}$$
i.e., the formula $act_{v}[\vX{},\vX{}']$ syntactically ensures the existence of a continuous transition.
Exploiting these relationships between our hybrid automata and the hybrid automata defined through the formul\ae\ 
$act_{v}[\vX{},\vX{}']$ and $trans_{e}[\vX{},\vX{}']$, we can reformulate the results presented in \cite{franzle99} in our framework.

\begin{definition}
Given a hybrid automaton $H=\langle\vX{}$, $\vX{}'$, $\Loc$, $\Edg$, $\InvSymb$,
   $\vfFunct{\cdot}$,  $\ActSymb$, $\ResetSymb\rangle$ we say that the hybrid automaton $\widetilde{H}=\langle\vX{}$, $\vX{}'$, $\Loc$, $\Edg$, $\widetilde{\InvSymb}$,
   $\widetilde{\vfFunct{\cdot}}$,  $\ActSymb$, $\ResetSymb\rangle$ is a \emph{disturbed variant} of $H$ if for each pair of states $\hstate{},\hstate{}'$ if $\hstate{}\actrans\hstate{}'$ in $H$, then $\hstate{}\actrans\hstate{}'$ in $\widetilde{H}$. 

Moreover, let $\edist$ be a distance over $\bbbr^{\dims{H}}$ and $\epsilon\in \bbbr_{>0}$.
$\widetilde{H}$ is a \emph{disturbance of noise level $\epsilon$ or more} if for each $s,s'$ such that $\edist(s,s')<\epsilon$ it holds that if $\state{v}{r}\actrans \state{v}{s}$ in $H$, then $\state{v}{r}\actrans \state{v}{s'}$ in $\widetilde{H}$.
\end{definition}

Intuitively, when there are no bifurcation behaviours, a small $\epsilon$ ensures that the dynamics of 
$\widetilde{H}$ are close to those of $H$.

In \cite{franzle99} it has been proved that in the case of bounded invariants there exists a finite computable index $i\in \mathbb N$ such that the reachability over $H$ can be over-approximated with reachability within $i$ discrete jumps over a disturbance of noise level $\epsilon$ or more of $H$. 

\begin{theorem}{\cite{franzle99}}
Let $\Theory$ be a decidable first-order theory.
Let $H,\widetilde{H}$ be $\Theory$ hybrid automata, with $\widetilde{H}$ disturbance of noise level $\epsilon$ or more of $H$ for some $\epsilon\in \bbbr_{>0}$. There exists $i$ such that for all pairs of states $\hstate{},\hstate{}'$ if $\hstate{}$ reaches $\hstate{}'$ in $H$, then $\hstate{}$ reaches $\hstate{}'$ in $\widetilde{H}$ within $i$ discrete transitions.
Moreover, $i$ can be effectively computed. 
\end{theorem}


Unfortunately, there are cases in which the over-approximation is always strict, no matter how small $\epsilon$ is. Such automata are called \emph{fragile}, in contrast with \emph{robust} automata, where if $\hstate{}$ does not reach $\hstate{}'$ in $H$, there exist $\epsilon \in \bbbr_{>0}$ and $\widetilde{H}$ disturbance of noise level $\epsilon$ such that $\hstate{}$ does not reach $\hstate{}'$ in $\widetilde{H}$. As a consequence, reachability is decidable over robust automata, while it is only semi-decidable over fragile automata. It is not possible to decide whether a hybrid automaton is robust or fragile. 
Intuitively, since real world systems are always subject to noise, a hybrid automaton is a reliable model of the real system only if it is robust. So, it is fundamental to develop and exploit design techniques which ensure robustness of the resulting hybrid automata.

\subsection{Approximate Bisimulations and Simulations}\label{sec:girard}

Since the 90's, simulation and bisimulation have been successfully used to investigate hybrid automata. 
However, due to the infinite precision required to relate different evolutions, these tools are able to 
remove neither the complexity nor the undecidability issues that may affect the analysis of the 
investigated model. 
The $\epsilon$-(bi)simulation relations~\cite{Girard:2007fk} relaxes these infinite precision requirements  by 
relating system evolutions whose maximal distance is less than a given $\epsilon$. 
This enables us to simplify both the dynamics and the resets of the investigated automaton. 
Moreover, provided an \emph{observation map} 
 $\obmap{\cdot}:\bbbr^{\dims{H}} \longrightarrow \bbbr^{d}$ that associates the internal status of an automaton $H$  
 to the values measurable by an external observer, $\epsilon$-(bi)simulations 
 allow to relate the ``visible'' behaviours of $H$ to 
 the behaviours of an automaton whose dimension is smaller than $\dims{H}$. 

Any pair of hybrid automata $H_1$ to $H_2$ 
related by an $\epsilon$-simulation must have the same discrete structure and share the same 
locations $\Loc$ and edges $\Edg$ by definition. 

\begin{definition}
Let $H_i=\langle\vX{i}$, $\vX{i}'$, $\Loc$, $\Edg$, $\InvSymb_i$,
   $\vfFunct{\cdot,i}$,  $\ActSymb_i$, $\ResetSymb_i\rangle$ be a hybrid automaton
    for each $i \in \{1,2\}$. 
   Moreover, let $\epsilon$ be in $\bbbr_{\geq 0}$. 
A relation $\mathcal{S}_{\epsilon} \subseteq  (\Loc \times \bbbr^{\dims{H_1}}) \times 
   (\Loc \times \bbbr^{\dims{H_2}})$ is an \emph{approximate simulation relation of 
  $H_1$ 
  by $H_2$ 
  of precision $\epsilon$}
  if, for all $\tuple{\state{\hloc{1}}{r_1}, \state{\hloc{2}}{r_2}} \in \mathcal{S}_{\epsilon}$: 
  \begin{enumerate}
  \item $\hloc{1}=\hloc{2}=\hloc{}$;
  \item $\|\obmap{r_1}_1-\obmap{r_2}_2\| \leq \epsilon $;
  \item if  $\state{\hloc{}}{r_1} \ctrans{t} \state{\hloc{}}{r_1'}$ in $H_1$, 
  there exists $r_2'$ s.t.~$\state{v}{r_2} \ctrans{t} \state{v}{r_2'}$ in $H_2$ and 
  $\tuple{\state{\hloc{}}{r_1'}, \state{\hloc{}}{r_2'}} \in \mathcal{S}_{\epsilon}$;
  \item  if  $\state{\hloc{}}{r_1} \dtrans{e} \state{\hloc{}'}{r_1'}$ in $H_1$, 
  there exists $r_2'$ s.t.~$\state{\hloc{}}{r_2} \dtrans{e} \state{\hloc{}'}{r_2'}$ in $H_2$ and 
  $\tuple{\state{\hloc{}'}{r_1'}, \state{\hloc{}'}{r_2'}} \in \mathcal{S}_{\epsilon}$.
   \end{enumerate}
\end{definition}

The automaton $H_2$ 
approximately simulates $H_1$ 
with precision $\epsilon$ if there exists an approximate simulation relation of 
  $H_1$ 
  by $H_2$ 
  of precision $\epsilon$. 
%
An approximate simulation relation $\mathcal{S}_{\epsilon}$ of 
  $H_1$ 
  by $H_2$ 
  of precision $\epsilon$ is an 
  \emph{approximate bisimulation relation between 
  $H_1$ 
  and $H_2$ 
  of precision $\epsilon$}
  if the relation $\mathcal{S}_{\epsilon}^{-1}=\{\tuple{\hstate{2},\hstate{1}} \, |\, \tuple{\hstate{1},\hstate{2}} \in \mathcal{S}_{\epsilon}\}$ 
  is an approximate simulation relation  of 
  $H_2$ 
  by $H_1$ 
  of precision $\epsilon$.

Many methods have been developed 
to automatically compute approximate simulation relations 
between systems such as 
constrained linear systems, 
autonomous nonlinear systems, and 
hybrid systems.

\subsection{$\epsilon$-Semantics}\label{sec:epsilon}

The undecidability of the reachability problem over hybrid automata having bounded invariants is a direct
consequence of the ability of characterizing dense regions of arbitrarily small size. 
As noticed in~\cite{DBLP:journals/deds/CasagrandePP09}, especially in the study of biological systems,
such ability may result misleading. As a matter of the fact, the continuous quantities used in hybrid automata
are very often abstractions of large, but discrete, quantities. 
In such cases, the ability of handling values with infinite precision is a model artifact rather than 
a real property of the original system. 

In order to discretize the continuous space, we introduce the concept of
\emph{$\epsilon$-sphere}. Given a set $\mathbb{S} \subseteq \mathbb{R}^n$, the
$\epsilon$-sphere $B(\mathbb{S},\epsilon)$ is the subset of $\mathbb{R}^n$
of points at distance less than $\epsilon$ from $\mathbb{S}$, i.e.,
$B(\mathbb{S},\epsilon) = \{ q \in \mathbb{R}^n \mid \exists p \in \mathbb{S}( \edist(p,q) < \epsilon) \}$,
where $\edist$ is a 
distance 
function 
over $\mathbb{R}^n$ (e.g., the standard euclidean distance). Moreover, given a hybrid
automaton $H$ and an initial set of points $\mathbb{I} \subseteq \bbbr^{\dims{H}}$, the
set of points reachable from the set $\mathbb{I}$ by $H$, denoted by $RSet_H(\mathbb{I})$, is
characterized by
$\ReachSet{H}{\mathbb{I}} = \bigcup_{i \in \mathbb{N}} \ReachSet[i]{H}{\mathbb{I}} = \lim_{i \rightarrow +\infty} \ReachSet[i]{H}{\mathbb{I}}$
where $\ReachSet[i]{H}{\mathbb{I}}$ is the set of points reachable from $\mathbb{I}$ in at
most $i$ discrete transitions.

\begin{theorem}[\cite{DBLP:journals/deds/CasagrandePP09}]
	Let $\mathcal{T}$ be a decidable first-order theory over reals and $H$ be a $\mathcal{T}$
	hybrid automaton with bounded invariants. If there exists $\epsilon \in \mathbb{R}_{> 0}$
	such that, for each $\mathbb{I} \subseteq \bbbr^{\dims{H}}$ and for each $i \in \mathbb{N}$,
	either $\ReachSet[i+1]{H}{\mathbb{I}} = \ReachSet[i]{H}{\mathbb{I}}$ or there exists an $a_i \in \mathbb{R}^{d(H)}$
	such that $B(\{a_i\},\epsilon) \subseteq \ReachSet[i+1]{H}{\mathbb{I}} \setminus \ReachSet[i]{H}{\mathbb{I}}$,
	then there exists $j \in \mathbb{N}$ such that $\ReachSet[i]{H}{\mathbb{I}} = \ReachSet[j]{H}{\mathbb{I}}$ and
	the reachability problem over $H$ is decidable.
\end{theorem}

This result finds applications when it makes no sense to distinguish measurements smaller than $\epsilon$.
Hence, since hybrid automata characterization is based on first-order fromul\ae, it seems reasonable to reinterpret the 
semantics of semi-algebraic automata by giving to each formula a semantics of \lq\lq dimension
of at least $\epsilon$\rq\rq. 
In~\cite{DBLP:journals/deds/CasagrandePP09} the authors introduce a new class of semantics, 
called \emph{$\epsilon$-semantics}, which guarantee the decidability of reachability
in the case of hybrid automata with bounded invariants.

%

\begin{definition}\label{gen:schema}
Let  $\Theory$ be a first-order theory and let $\epsilon \in \bbbr_{>0}$.
For each formula $\psi$ on $\Theory$ let $\epssem{ \psi }\subseteq \bbbr^d$, where $d$ is the number of free variables of $\psi$, be such that:
\begin{itemize}
 \item[($\epsilon$)] either $\epssem{ \psi }=\emptyset$ or there exists $p\in \mathbb R^d$ such that $\eball{\{p\}}\subseteq \epssem{ \psi }$;
 \end{itemize}
\begin{minipage}[b]{0.5\linewidth}
\begin{itemize}
\item[($\cap$)] $\epssem{ \phi\land \varphi} \subseteq \epssem{ \phi} \cap \epssem{\varphi}$;
\item[($\cup$)] $\epssem{ \phi\lor \varphi} = \epssem{ \phi} \cup \epssem{\varphi}$;
\end{itemize}
\end{minipage}
\begin{minipage}[b]{0.5\linewidth}
\begin{itemize}
\item[($\forall$)] $\epssem{ \forall \X{} \NotBoundIn{\psi}{\X{},\vX{}} } = 
\epssem{\bigwedge_{r \in \bbbr} \VarSSub{\psi}{r,\vX{}}}$;
\item[($\exists$)] $\epssem{ \exists \X{} \NotBoundIn{\psi}{\X{},\vX{}} } = 
\epssem{\bigvee_{r \in \bbbr} \VarSSub{\psi}{r,\vX{}} }$;
\end{itemize}
\end{minipage}
\begin{itemize}
\item[($\neg$)] $\epssem{ \psi }\cap \epssem{ \neg \psi }=\emptyset$.
\end{itemize}
Any semantics satisfying the above conditions is said to be an 
\emph{$\epsilon$-semantics for $\Theory$}. 
\end{definition}

Let us notice that 
no $\epsilon$-semantics can over-approximate 
the standard semantics. Infact, 
for any theory $\Theory$, if $\stdsem{\phi} \subset \epssem{\phi}$, 
where $\stdsem{\cdot}$ is the semantics 
associated to $\Theory$, then $\stdsem{\neg \phi} \supset \epssem{\neg \phi}$ due to the ($\neg$)-rule.

\begin{example}[The sphere semantics]
\label{spheresemantics}
Let $\Theory$ be a first-order theory over the reals and let $\epsilon >0$. 
The \emph{sphere semantics} of $\psi$, $\sphsem{\psi}$, is defined by structural induction on $\psi$ as 
follows:

\begin{minipage}[b]{0.5\linewidth}
\begin{itemize}
\item $\sphsem{t_1 \circ t_2}\defeq\eball{\stdsem{t_1\circ t_2}}$, for $\circ \in \{=,<\}$;
\item $\sphsem{ \psi_1\wedge \psi_2}\defeq\bigcup_{\eball{\{p\}}\subseteq \sphsem{ \psi_1}\cap \sphsem{\psi_2}}\eball{\{p\}}$;
\item $\sphsem{ \psi_1\vee \psi_2}\defeq\sphsem{ \psi_1}\cup \sphsem{ \psi_2}$;
 \end{itemize}
 \end{minipage}
\begin{minipage}[b]{0.5\linewidth}
\begin{itemize}
\item $\sphsem{ \forall \X{} \VarSSub{\psi}{\X{},\vX{}} } \defeq
\sphsem{\bigwedge_{r \in \bbbr}  \VarSSub{\psi}{r,\vX{}}}$;
\item $\sphsem{ \exists \X{} \VarSSub{\psi}{\X{},\vX{}{}} } \defeq
\sphsem{\bigvee_{r \in \bbbr} \VarSSub{\psi}{r,\vX{}{}} }$;
\item $\sphsem{ \neg \psi }\defeq \bigcup_{\eball{\{p\}}\cap \sphsem{ \psi}=\emptyset}\eball{\{p\}}$.
 \end{itemize}
\end{minipage}
\end{example}

	
$\epsilon$-semantics are exploited in the reachability algorithm defined in~\cite{DBLP:journals/deds/CasagrandePP09}.
Intuitively, the algorithm computes reachability incrementing the number
of allowed discrete transitions at each iteration. New reachable sets of points are computed until
they became too small to be captured by the $\epsilon$-semantics.	
In the case of hybrid automata with bounded invariants, it 
always terminates. 
Finally, notice that replacing the $\epsilon$-semantics with the standard one,
the above algorithm could not terminate even with bounded invariants, due to Zeno behaviours.

To conclude, let us notice that, by Taylor's approximation, for any $\epsilon$ and any differentiable function $f(t)$, 
if we fix a time horizon $t_{h}$, we can approximate $f(t)$ 
by a polynomial $p(t)$ such that $\|f(t)-p(t)\|<\epsilon$ for all 
$t\in [0,t_{h}]$. Hence, there are $\epsilon$-semantics that are not able to distinguish 
$p(t)$ and $f(t)$ and, according to such $\epsilon$-semantics, we can use $p(t)$ in place of $f(t)$ for all $t \in [0,t_{h}]$.
It follows that, adopting an opportune $\epsilon$-semantics, the Tarski's theory (i.e., the first order theory of 
the inequalities over the reals) is enough to represent any differentiable function 
$f(t)$.

\section{Approximated Analysis over Neural Oscillator}\label{sec:analysis}
In this section we try to understand what happens when we apply the approximation techniques described in Section~\ref{sec:approx} to our neural oscillator
hybrid model $H_{\tilde{f}}$ presented in Section~\ref{sec:model}. 
\subsection{$\epsilon$-disturbance}\label{subsec:franzle_analysis}
The automaton $H_{\tilde{f}}$ presents two main behaviours: $\tuple{0,0}$ is an unstable equilibrium;
each starting point different from $\tuple{0,0}$ reaches the limit cycle.
For this reason we can prove that $H_{\tilde{f}}$ is fragile. As a matter of fact, if $\widetilde{H_{\tilde{f}}}$ is
a disturbance of noise level $\epsilon$ of $H_{\tilde{f}}$, then $(0,0)$ in $\widetilde{H_{\tilde{f}}}$ reaches points 
different from $\tuple{0,0}$, while in $H_{\tilde{f}}$ it does not. In other words, if we consider backward reachability, $\tuple{0,0}$ is backward
reachable in $\widetilde{H_{\tilde{f}}}$  from a region $R\neq \{(0,0)\}$, while in $H_{\tilde{f}}$ it is backward reachable from
$\{\tuple{0,0}\}$.
This is not due to the fact that $\tuple{0,0}$ is unstable, but to the presence of two limit behaviours over a connected region.
We recall that in a piecewise hybrid automaton the invariants are connected disjoint regions whose union is connected and the resets are identities, i.e., 
the trajectories are continuos. We use the term \emph{limit behaviour} of a hybrid automaton to denote both equilibria and limit 
cycles.
\begin{theorem}
Let $H$ be a piecewise hybrid automaton presenting at least two different limit behaviours. 
If from each point there is a unique possible evolution, then $H$ is fragile.
\end{theorem}
%
The above result points out that there are systems for which it is not possible to define robust models.
In~\cite{DBLP:conf/tamc/Ratschan10} a model is said to be safe only if it remains safe under small disturbances. In this 
terms our result shows that
there are systems which do not admit a safe model. This does not means that they are not interesting or that we need to 
remove
some of their behaviours. This simply means that such systems have to be studied applying some form of disturbance or 
approximation. As the matter of facts, if we study them by applying standard semantics, we define a precise border between 
the points reaching different
behaviours. Such precise border is not realistic. 
 
\subsection{$\epsilon$-(bi)simulations}\label{subsec:girard_analysis}

The automaton $H_{\tilde{f}}$ has both an unstable equilibrium in $\tuple{0,0}$ and a single limit 
cycle encompassing the origin of the axes. 
Because of that we are guaranteed that,  during its evolutions, 
$H_{\tilde{f}}$ decreases the 
distance of its state from the limit cycle 
regardless of the starting state $s\neq \state{v_0}{\tuple{0,0}}$. 
Since all the differential equations defining the dynamics of $H_{\tilde{f}}$ are continuous,  
we can define an $\epsilon$-simulation between states whose distance 
from the limit cycle is smaller than $\epsilon$. 
%
This enables us to both approximate the non-linear differential System (\ref{eq:origsystem})
 with a linear differential system and reduce  the complexity of 
the analysis. 

However, if $d$ is the maximum 
Euclidean distance between $\tuple{0,0}$ and the cycle limit,  
no $\epsilon$-(bi)si\-mu\-la\-tion, 
with $\epsilon<d$,   
can relate $\state{v_0}{\tuple{0,0}}$ with any other state 
of $H_{\tilde{f}}$. As a matter of fact, the points belonging to any 
neighborhood of $\tuple{0,0}$ eventually converge to 
the limit cycle.
It follows that $\state{v_0}{\tuple{0,0}}$ is 
a singularity of the model and, despite the original system 
always reaches a periodic evolution, any approximation of the proposed 
model by mean of $\epsilon$-(bi)simulation does not manifest this 
property. 

\subsection{$\epsilon$-semantics}\label{subsec:epsilon_analysis}
In order to exploit $\epsilon$-semantics for the study of $H_{\tilde{f}}$ the
first step we have to perform is that of approximating through polynomials
the solutions of the differential equations defining the semantics.
This can be done, for instance, by using Taylor polynomials or more sophisticated
numerical integration techniques.
We do this in the next section, where we also apply cylindrical algebraic decomposition
tools to automatically prove properties of our model.
Here instead we try to infer some general results about the use of $\epsilon$-semantics on
$H_{\tilde{f}}$.

$H_{\tilde{f}}$ has an unstable equilibrium in $\tuple{0,0}$. This means that
$\tuple{0,0}$ reaches $\{\tuple{0,0}\}$. However, when we compute the set of points
reachable from $\tuple{0,0}$ through an $\epsilon$-semantics we get either 
the empty set or a set having diameter at least $\epsilon$.
In particular, if our $\epsilon$-semantics under-approximates the standard
one (i.e., if $\stdsem{\phi} \supset \epssem{\phi}$, 
where $\stdsem{\cdot}$ is the semantics 
associated to chosen theory, for each formula $\phi$), then we get the empty set. Otherwise, both cases are possible, depending
on the $\epsilon$-semantics. For instance, in the case of sphere semantics,
no matter how we approximate the dynamics, we get that $\tuple{0,0}$ reaches
a set having diameter at least $\epsilon$.
Similarly, unless we use an under-approximation $\epsilon$-semantics or 
some unusual metrics, the limit cycle is transformed into a limit flow tube.
This means that, if we consider a point on the limit cycle and we compute the
set of points reachable from such point, then we do not only obtain the limit cycle, but
at least a flow tube which includes the limit cycle. We will see some more
details on this in the case of sphere semantics in Section~\ref{sec:computing}.
All the other points, again, will reach either the empty set or a set having diameter
at least $\epsilon$. The result we would expect in this second case is that each point in the
space reaches the flow tube including the limit cycle. We will see that this is true in the case
of sphere semantics, even when we use the simplest Taylor polynomials of degree one. 

These considerations already allow us to point out that sphere semantics better reflects 
the real system behaviour than the standard one. 

\section{Computing Sphere Semantics}\label{sec:computing}
In this section we show how sphere semantics can be computed exploiting tools for cylindrical algebraic decomposition.
In particular, we introduce a translation from sphere semantics to standard semantics.
Then, we apply the translation to study the ``sphere'' behaviour of our neural oscillator example.

\subsection{A translation into standard semantics}

If $\Theory$ is a first-order theory and $\edist$ is a distance definable in $\Theory$, then 
the sphere semantics of any formula in $\Theory$ is $\Theory$-definable in the standard semantics, i.e., 
for any formula $\NotBoundIn{\varphi}{\vX{}} \in \Theory$  
we can compute a formula $\NotBoundIn{\tostdsem{\varphi}}{\vX{}} \in \Theory$ such that 
$\sphsem{\NotBoundIn{\varphi}{\vX{}} } = \stdsem{ \NotBoundIn{\tostdsem{\varphi}}{\vX{}}  }$ for all $\epsilon \in \bbbr_{>0}$. 

In order to achieve this goal, we need to distinguish two kind 
of variables: the variables of the original formula (named $\W{}$, $\W{i}$, $\vW{}$ and $\vW{i}$), whose evaluations follow 
the rules of the sphere semantics, and the auxiliary variables (named $\Y{}$, $\Y{i}$, $\vY{}$ and $\vY{i}$)
that will be introduced to encode the sphere semantics into the standard one. From the point of view 
of the sphere semantics the later can seen as symbolic constants, even if they will be quantified in the formula 
$\tostdsem{\varphi}$. In particular, we will use them to 
characterize sets of the form $\sphsem{\bigwedge_{r \in \bbbr} \VarSSub{\varphi}{r,\vW{}}}$ and 
$\sphsem{\bigvee_{r \in \bbbr} \VarSSub{\varphi}{r,\vW{}}}$ in the standard semantics. 

\begin{definition}[Translation]\label{def:tostdsem}
Let $\Theory$ be a first-order theory over the reals, $\NotBoundIn{\varphi}{\vY,\vW{}}$ be any first-order formula $\Theory$-definable, and $\epsilon \in  \mathbb{R}_{> 0}$.
We define $\NotBoundIn{\tostdsem{\varphi}}{\vY,\vW{}}$ by structural induction on $\NotBoundIn{\varphi}{\vY,\vW{}}$ itself.  
\begin{enumerate}
\item\label{trans:atomic} $\tostdsem{\NotBoundIn{(t_1 \circ t_2)}{\vY{},\vW{}}} \defeq \exists \vW{0}(\VarSSub{(t_1 \circ t_2)}{\vY{},\vW{0}} \wedge \edist(\vW{0},\vW{}) < \epsilon)$, for $\circ \in \{=,<\}$;
\item\label{trans:disjunction} 
$\tostdsem{\phi\lor\psi} \defeq %
\tostdsem{\phi} \lor \tostdsem{\psi}$;
\item\label{trans:conj} $\tostdsem{\NotBoundIn{\phi}{\vY{},\vW{}} \land \NotBoundIn{\psi}{\vY{},\vW{}}} \defeq 
\exists \vW{0} (\forall \vW{1} (\edist(\vW{0},\vW{1}) < \epsilon \limply \VarSSub{(\tostdsem{\phi}\land \tostdsem{\psi})}{\vY{},\vW{1}})\land \edist(\vW{0},\vW{}) < \epsilon)$;
\item\label{trans:for} $\tostdsem{\forall \W{} \NotBoundIn{\phi}{\vY{},\W{},\vW{}}} \defeq 
\exists \vW{0} ( \forall \vW{1} (\edist(\vW{0},\vW{1}) < \epsilon \limply \forall \Y{} \tostdsem{\VarSSub{\phi}{\vY{},\Y{},\vW{1}}})\land \edist(\vW{0},\vW{}) < \epsilon)$;
\item\label{trans:exists} $\tostdsem{\exists \W{} \NotBoundIn{\phi}{\vY{},\W{},\vW{}}} \defeq  \exists \Y{} \tostdsem{\VarSSub{\phi}{\vY{},\Y{},\vW{1}}}$;
\item\label{trans:neg} $\tostdsem{\NotBoundIn{\neg \phi}{\vY{},\vW{}}} \defeq 
\exists \vW{0} ( \forall \vW{1} (\edist(\vW{0},\vW{1}) < \epsilon \limply \neg \tostdsem{\VarSSub{\phi}{\vY{},\vW{1}}})\land \edist(\vW{0},\vW{}) < \epsilon)$.
\end{enumerate}

\end{definition}

\begin{theorem}[Semantics Equivalence]
Let $\Theory$ be any first-order theory and $\edist$ be a  
$\Theory$-definable distance. 
The sphere semantics $\sphsem{.}$ of $\Theory$ is $\Theory$-definable in the standard semantics and,  
in particular, 
$\sphsem{\NotBoundIn{\varphi}{\vX{}} } = \stdsem{ \NotBoundIn{\tostdsem{\varphi}}{\vX{}}  }$ 
for any formula $\NotBoundIn{\varphi}{\vX{}} \in \Theory$ and all $\epsilon \in \bbbr_{>0}$.
\end{theorem}

\noindent Since $\NotBoundIn{\tostdsem{\varphi}}{\vY,\vW{}}$ is 
definable in Tarski theory, and
	this theory is decidable, the satiability of $\varphi$ in $\sphsem{}$ is decidable.


Let us notice that the application of the translation in Definition~\ref{def:tostdsem} to a formula, increases 
the evaluation complexity of such formula with respect to its untranslated version. This is mainly due to
the possible introduction of quantifier operator alternations.

\subsection{Experimental Results on the Neural Oscillator}

Let us consider the hybrid automaton $H_{\tilde{f}}$ described in Section \ref{sec:model} for modeling a neural oscillator. 
We intend to study its behaviour through sphere semantics, exploiting cylindrical algebraic decomposition tools to 
automatically compute it.

\begin{figure}[!h]
\begin{center}
\includegraphics[width=0.67\textwidth]{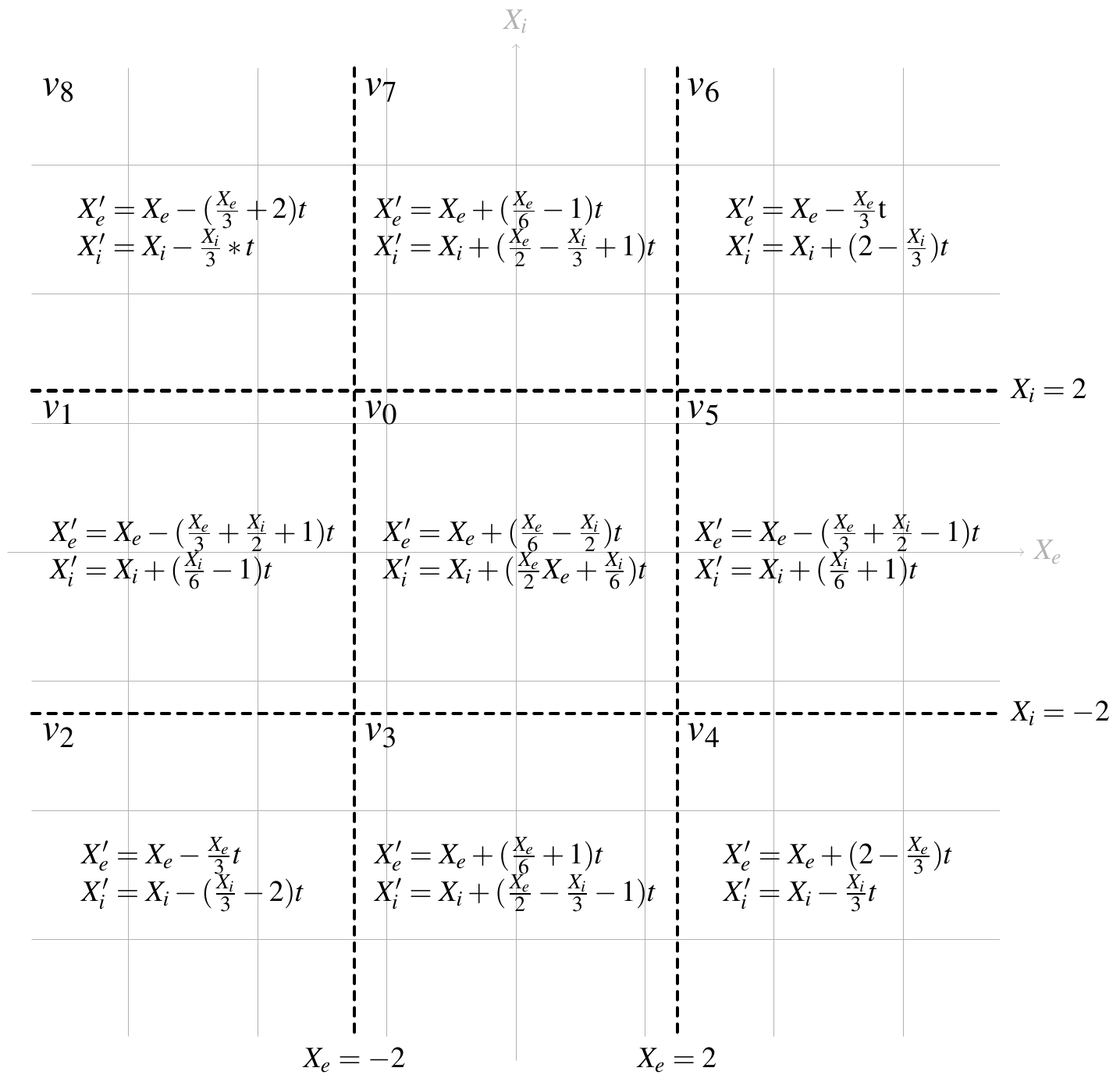}
\caption{The piecewise hybrid automaton $H'_{\tilde{f}}$ associated to the function $f_{2}(3,1)$.}\label{fig:taylorapproxfunct}
\end{center}
\end{figure}

As we noticed in Section~\ref{sec:approx},  
any differentiable dynamics can be exactly represented, in terms of an opportune $\epsilon$-semantics, by a semi-algebraic function. 
In particular, we can replace the dynamics of $H_{\tilde{f}}$ by the corresponding 
Taylor polynomials up to a certain degree which depends on $\epsilon$.
In this paper, we decided to model the 
automaton dynamics by using their first-degree Taylor polynomials 
and we obtained 
%
the automaton $H'_{\tilde{f}}$ depicted in Figure~\ref{fig:taylorapproxfunct}.
In order to keep the presentation simple, in this section we fix the parameters as follows 
$\tau=3$, $\lambda=1$, $\alpha=2$. Hence, the activations correspond to the axis $\X{i}=\pm 2$ and $\X{e}=\pm 2$.  

A simulation of $H'_{\tilde{f}}$ is presented in Figure \ref{fig:diamond_cycle}: 
a limit cycle is still present, 
but it has a diamond-like shape. We are interested in studying this limit cycle. In particular, we are interested in proving, exploiting tools for symbolic computation, that if we apply sphere semantics, each point in the space reaches a bounded region 
which includes the limit cycle. 
In this example our automata have unbounded invariants, hence the termination of 
sphere semantics reachability algorithm is not guaranteed.

We start computing the intersections of the limit cycle with the activation regions. 
Consider for instance the intersection $Q_0=\tuple{x_{Q_0},2}$ of the limit cycle with $\X{i}=2$ and 
$\X{e}>0$.
We have that $x_{Q_0}$ is the unique solution of the equation which describes the intersection of the
diamond-like limit cycle with $\X{i}=2$. Similarly, consider point $Q_1=\tuple{2,y_{Q_1}}$
that in turn corresponds to the intersection of the limit cycle with $\X{e}=2$ and $\X{i}>0$. 
We effectively calculated all these intersections by using the computer algebra system
 {\tt Maxima}. 
So, for instance, we get $x_{Q_0}=\frac{3526\,\sqrt{17}+14538}{495\,\sqrt{17}+2041}$ and 
$y_{Q_1}=\frac{190\,\sqrt{17}+786}{39\,\sqrt{17}+161}$.
The points 
$Q_0$ and $Q_1$ satisfy the activation formul\ae~wich regulate 
the discrete transitions between locations $\hloc{6}$ and $\hloc{5}$, and locations $\hloc{6}$ and $\hloc{7}$, 
respectively. Let us now consider a point $P_0$ located on $\X{i}=2$, but which is such
that its distance $d_0$ from $Q_0$ is at least $2\epsilon$, i.e., $P_0=\tuple{x_{P_0},2}$ and
$\edist(Q_0,P_0) = d_0 > 2\epsilon$. Consider now any point $P_1$ on $\X{e}=2$ resulting 
from the sphere semantics evaluation of the continuous evolution which starts in $P_0$
inside location $\hloc{6}$. Thus, let denote with $d_1$ the distance between such $P_1$ and $Q_1$.

If we could prove that $d_1$ is always smaller than $d_0$, then we would be able to conclude that all the points 
which start from a distance of at least $2\epsilon$ from the limit cycle converge to a flow tube  
having diameter $2\epsilon$ that includes the limit cycle. Of course, to obtain such conclusion, we need to prove 
this property on all locations. 

We can formalize this concept through a first-order formula. We denote with $r$ and $s$ the straight lines $\X{i}=2$ 
and $\X{e}=2$, respectively, and with the notation $Q_0 \in r \cap C \cap \X{e}>0$ the membership of $Q_0$ to the 
intersection of straight line $r$ with limit cycle $C$ and positive $\X{e}$ semi-plane. Moreover, with the notation
$\sphsem{ P_0 \rightarrow_C P_1 }$ we denote the continuous transition from point $P_0$ to point 
$P_1$ performed exploiting sphere semantics. Thus, our desired property can be expressed as:
\begin{equation}\label{eq:epsilon_conv}
\begin{split}
\forall Q_0 Q_1 \forall P_0 P_1 \big( ( & Q_0 \in r \cap C \cap \X{e}>0 \wedge Q_1 \in s \cap C \cap \X{i}>0 \wedge P_0 \in r \cap \X{e}>0 \wedge  \\
		&P_1 \in s \cap \X{e}>0 \wedge \edist(Q_0,P_0) > 2\epsilon \wedge \sphsem{ P_0 \rightarrow_C P_1 }) \limply \edist(Q_1,P_1) < \edist(Q_0,P_0) \big)
\end{split}
\end{equation}
stating the convergence to the limit flow tube in location $\hloc{6}$. Such property can be 
easily rewritten for each location of the hybrid automaton, changing the roles of activation border lines 
$r$ and $s$.

\begin{figure}[!ht]
 \centering
   \includegraphics[width=0.50\textwidth]{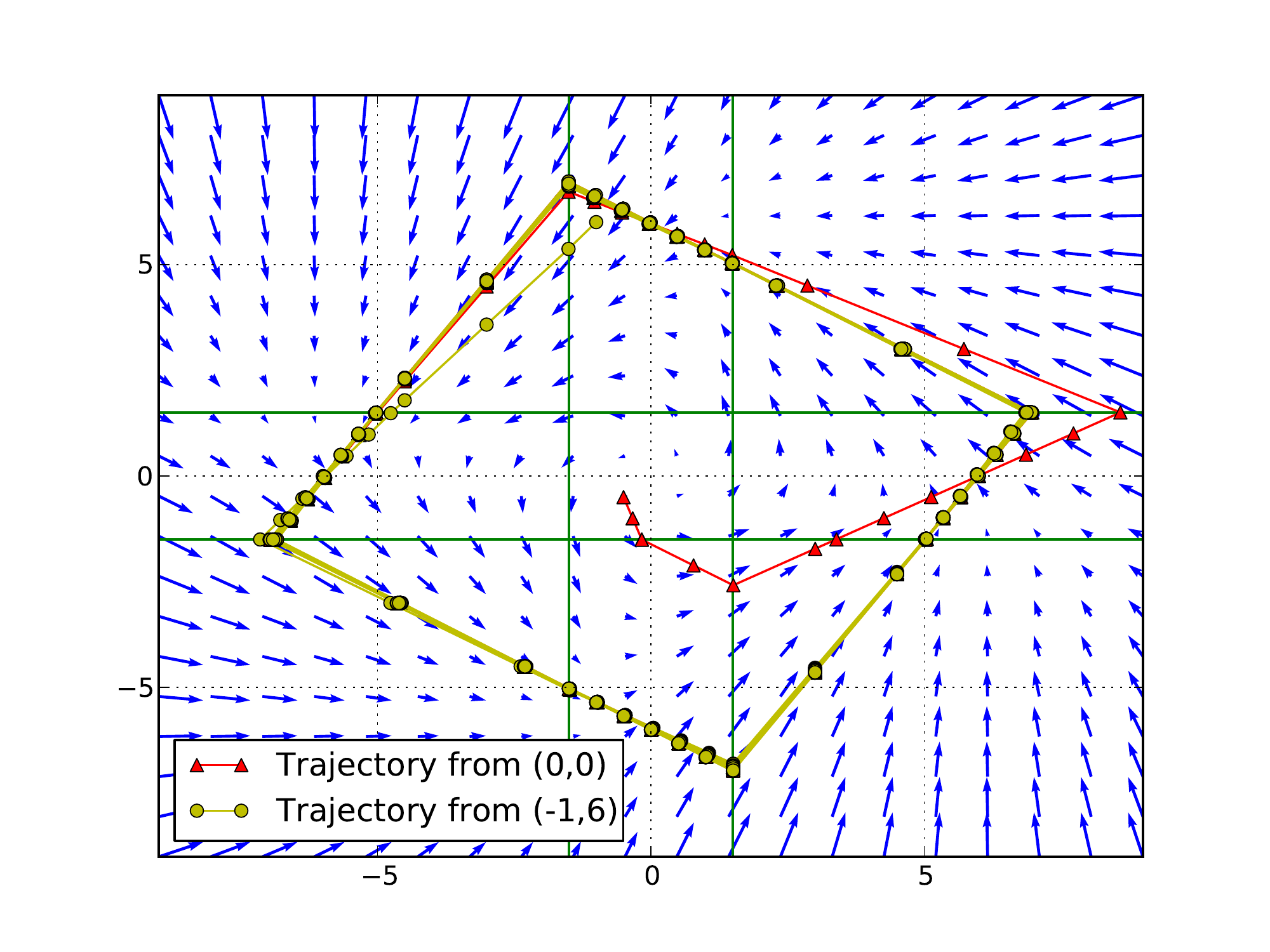}
    \caption{Two evolutions of the first degree approximation of the model proposed in Section~\ref{sec:model}.}\label{fig:newtonev}
 \end{figure}
 
We automatically expanded such formula by using a Perl script that implements the Definition~\ref{def:tostdsem} to translate sphere semantics into the standard one. 
In particular, $\sphsem{ P_0 \rightarrow_C P_1 }$, in the case of location $\hloc{6}$ becomes 
\begin{equation}\label{eq:full_formula}
\NotBoundIn{\psi}{\vX{}} \defeq \exists \Tvar{} (\exists \vX{0} ( 
\forall \vX{1} (  \edist{(\vX{0},\vX{1})}< \epsilon \limply (\Tvar{}>0 \land 
\VarSSub{\phi}{\vX{1},\Tvar{}}))) \land  \edist{(\vX{0},\vX{})}< \epsilon), 
\end{equation}
where 
\begin{equation}
\begin{split}
\NotBoundIn{\phi}{\vX{},\Tvar{}} \defeq \exists \vX{0} (\forall \vX{1} ( \edist{(\vX{0},\vX{1})}< \epsilon &\limply 
( \exists \vX{2} (\VarSSub{\Pi_1}{\vX{2},\Tvar{}} \land \edist{(\vX{2},\vX{1})}< \epsilon ) \land  \\
&\exists \vX{2} (\VarSSub{\Pi_2}{\vX{2},\Tvar{}} \land \edist{(\vX{2},\vX{1})}<\epsilon))) \land \edist{(\vX{0},\vX{})}<\epsilon) ,
\end{split}
\end{equation} 
$\NotBoundIn{\Pi_1}{\X{0},\Y{0},\X{1},\Y{1},\Tvar{}} \defeq 6*\X{1}=6*\X{0}+(\X{0}-3*\Y{0})*\Tvar{}$, and 
$\NotBoundIn{\Pi_2}{\X{0},\Y{0},\X{1},\Y{1},\Tvar{}} \defeq 6*\Y{1}=6*\Y{0}+(\Y{0}+3*\X{0})*\Tvar{}$.
%
However, we notice that, since $\Pi_1$ and $\Pi_2$ are closed and convex, $\phi$ can be simplified as: 
\begin{equation}
\NotBoundIn{\phi}{\vX{},\Tvar{}} \equiv \exists \vX{0} (\VarSSub{\Pi_1}{\vX{0},\Tvar{}} \land \VarSSub{\Pi_2}{\vX{0},\Tvar{}} \land \edist{(\vX{0},\vX{})}<\epsilon).
\end{equation}
 Similarly, $\psi$ becomes:
 \begin{equation}
\NotBoundIn{\psi}{\vX{}} \equiv \exists \Tvar{} (\Tvar{}>0 \land \exists \vX{0} (\VarSSub{\Pi_1}{\vX{0},\Tvar{}}
 \land \VarSSub{\Pi_2}{\vX{0},\Tvar{}} \land \edist{(\vX{0},\vX{})}<\epsilon)).
\end{equation}
So we plugged this last formula in Formula~\ref{eq:epsilon_conv} and used {\tt REDLOG}~\cite{Dolzmann:1997:RCA:261320.261324} to test it. 
The formula turns out to be true (the result is computed within few seconds), proving our conjectures. 

Notice that we used {\tt Maxima} to compute the exact coordinates of the points on the limit cycles since that computation 
does not require quantifier elimination. However, we could have used {\tt REDLOG}. 

As far as $\tuple{0,0}$ is concerned it is immediate to prove through a first-order formula that it reaches points different from 
itself and, hence, it reaches the limit flow tube. 

Other interesting properties that automatically verified express, for instance, the fact that applying the sphere semantics there are points that cross 
the limit cycle (in both directions). This is quite natural since points closer than $\epsilon$ to the limit cycle get expanded and cross it.

\section{Conclusions}\label{sec:conclusions}


In this paper we have modeled a neural oscillator constructing a hybrid automaton
whose components derive from the approximation of the continuous model presented in~\cite{Tonnelier19991213}.
We have analyzed its behaviours considering the application of some approximation techniques for the introduction of noise, as
already advocated in ~\cite{Tonnelier19991213}.
In particular, we focused on the approach based on the $\epsilon$-semantics.

The simulation based on the application of the $\epsilon$-semantics has revealed the any point which begins its 
evolution from a distance of at least $2\epsilon$ from the limit cycle, converges to a flow tube 
which possesses a diameter equal to $2\epsilon$ and that includes the limit cycle. 
Due to size of the formul\ae\ which compose the hybrid automaton and the growth of such formal\ae\ introduced 
by the translation of the $\epsilon$-semantics evaluations, a direct computation of the reachable set would have
high complexity and eventually returns results of difficult interpretation.
For this reason, we have reformulated the problem 
in form of a closed property which guarantees the convergence of any point towards the limit cycle of the modeled system.

During the construction of the formula that describes the convergence to the 
limit cycle, some steps of simplification of the formul\ae\ have been applied. In particular, 
we have reduced the complexities of translated formul\ae, 
relying on the convexity of the sets characterized by some of their subformul\ae.
An interesting aspect to investigate is whether these simplification steps 
can be automatically performed. 

As future work, in order to analyze the behaviour of a group of neural oscillators, we plan 
to combine several hybrid automata and to study their evolutions always adopting
$\epsilon$-semantics. 

\bibliographystyle{eptcs}
\bibliography{hybridbib}
\end{document}